\documentclass[doublecol]{epl2} 
\title{Structural domain and spin ordering induced  glassy magnetic phase in single layered manganite Pr$_{0.22}$Sr$_{1.78}$MnO$_4$}
\shorttitle{Glassy magnetic phase in single layered manganite  Pr$_{0.22}$Sr$_{1.78}$MnO$_4$}
\author{S. Chattopadhyay\inst{1} \and S. Giri\inst{1} \and S. Majumdar\inst{1}\thanks{E-mail: \email{sspsm2@iacs.res.in}}}
\shortauthor{S. Chattopadhyay \etal}
\institute{\inst{1} Department of Solid State Physics, Indian Association for the Cultivation of Science, Jadavpur, Kolkata 700032, India}

\pacs{75.25.Dk}{Orbital, charge, and other orders, including coupling of these orders.}

\abstract{The single layered manganite  Pr$_{0.22}$Sr$_{1.78}$MnO$_4$ undergoes  structural transition from  high temperature tetragonal phase to low temperature orthorhombic phase below room temperature. The orthorhombic phase was reported to have two structural variants with slightly different lattice parameters and  Mn-3$d$ levels  show orbital ordering within both the variants, albeit having mutually perpendicular ordering axis. In addition to orbital ordering, the orthorhombic variants also order antiferromagnetically  with different N\'eel temperatures. Our magnetic investigation on the polycrystalline sample of Pr$_{0.22}$Sr$_{1.78}$MnO$_4$ shows large thermal hysteresis indicating the first order nature of the tetragonal to orthorhombic transition. We observe magnetic memory, large relaxation, frequency dependent ac susceptbility  and aging effects at low temperature, which indicate spin glass like magnetic ground state in the sample. The  glassy magnetic state presumably arises from the interfacial frustration of orthorhombic domains with orbital and spin orderings playing crucial role toward the competing magnetic interactions.} 

\begin{document}

\maketitle

One of the intriguing aspects of AMnO$_3$ type perovskite manganites (A = alkaline earth or rare-earth element) is the frequent observation of glassy magnetic state at low temperature, which is believed to arise from the competing interactions between finite size ferromagnetic (FM) and antiferromagnetic (AFM) clusters~\cite{dag}. AMnO$_3$ compounds actually belong to a wider class of materials known as Ruddlesden-Popper (RP) phase with general formula A$_{n+1}$Mn$_n$O$_{3n+1}$, and they may be thought of as the $n = \infty$ member in the RP phase. There also exist several other manganites in the RP phase with fascinating crystal structures. The lowest possible value of $n$ is 1 and this corresponds to A$_2$MnO$_4$ type single layered manganites~\cite{jea,moh,bao,ima,coe,kim,yu}. These compounds crystallize in a tetragonal K$_2$NiF$_4$ type structure where  MnO$_2$ layers remain separated from one another through A$^{2+}$ ions and construct a quasi-two-dimensional (2D)  lattice structure.

\begin{figure}[t]
\vskip 0.1 cm
\centering
\includegraphics[width = 7 cm]{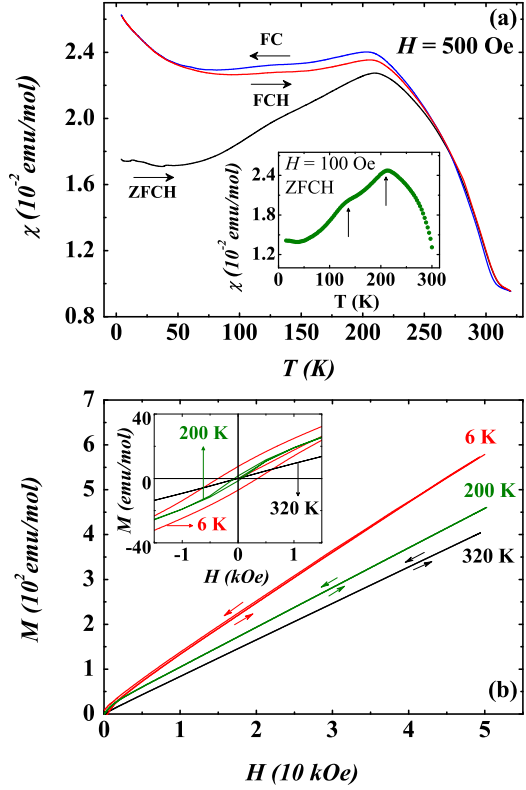}
\caption {(a) shows the dc magnetic susceptibility measured with 500 Oe of field  in the zero-field-cooled heating (ZFCH), filed-cooled heating (FCH) and field-cooling (FC) protocols. The inset shows the ZFCH curve for 100 Oe of applied field. (b) shows isothermal magnetization as a function of applied field ($H$) at three different temperatures. The inset shows the enlarged view of the curves in the low field region. }
\end{figure}
\begin{figure}[t]
\vskip 0.1 cm
\centering
\includegraphics[width = 7 cm]{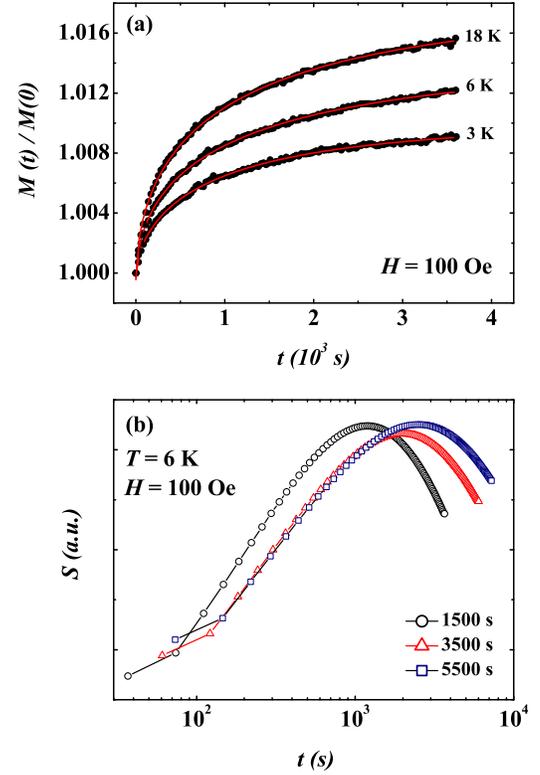}
\caption {(a) shows the isothermal time dependence of magnetization [$M(t)]$ measured at three different temperatures. Here the  data have been normalized by the initial value of the magnetization [$M(0)]$. The solid lines  are fit to the data with stretched exponential form of relaxation. (b) shows the magnetic viscosity ($S = \frac{1}{H}\frac{\partial M}{\partial (\ln t)} $) at 6 K in the ZFC state as a function of time. Here three different curves correspond to measurement in 100 Oe  after initial  wait at zero field for $t_w$ = 1500, 3500 and 5500 s respectively. }
\end{figure}

\par
Similar to AMnO$_3$ compounds, doping with  rare-earth ions at the $A$ site of A$_2$MnO$_4$ produces mixed valency where both Mn$^{3+}$ and  Mn$^{4+}$ ions can exist. However, unlike perovskite system, single layered  manganites (SLMs) do not show FM-metallic ground state and colossal magnetoresistance is absent. Rather, in many situations, they undergo charge ordering /orbital ordering (CO/OO) which ultimately favors an AFM ground state~\cite{kim,ste,tok}. Due to the absence of finite FM region or cluster,  SLMs are  unlikely to have a cluster glass like state. However, it should be kept in mind that FM bonds between Mn atoms are actually present in SLMs, which along with super exchange mediated AFM bonds  can give rise to long range C or CE type AFM structure. Many SLMs show complex interplay between orbital, magnetic, structural and charge degrees of freedom,  and it is pertinent to investigate the magnetic ground state of such materials, particularly in  presence of coexisting FM and AFM bondings.

\par

In this work, we chose a suitable SLM with  nominal composition Pr$_{0.22}$Sr$_{1.78}$MnO$_4$ (PSMO) for such magnetic investigations.  The series of compounds, Pr$_{1-x}$Sr$_{1+x}$MnO$_4$ (0.75 $\leq$ $x$ $\leq$ 0.9) are reported to undergo a structural phase transition from tetragonal (TG) to orthorhombic (OR) phase with lowering of $T$. {\it Interestingly, there are two orthorhombic variants OR1 and OR2 with unequal $b$/$a$ ratios ($a$ and $b$ are the lattice parameters). Both the variants undergo orbital ordering, although with mutually orthogonal orientations.} In the OR1 phase, OO takes place in the $d$(3$x^{2}$ - $r^{2}$) orbital of $Mn^{3+}$ ions, whereas, it is $d$(3$y^{2}$ - $r^{2}$) orbital that stabilizes in the OR2 phase. The structural phase transitions, {\it i.e.} TG to OR1 and TG to OR2 also occur at two different temperatures. Transmission electron microscopy reveals the presence of alternating arrays of orbitally ordered OR1 and OR2 domains~\cite{nor} at low temperature. Similar orthorhombic domain structure is also observed in few other compositions among SLMs with general formula R$_{1-x}$Sr$_{1+x}$MnO$_4$ (R = La, Nd)~\cite{nor2,nor3,nor4,nag}. With further lowering in temperature ($T$), OR1 phase in PSMO undergoes a C-AFM transition at $T_{N1}$ where orbitally ordered OR2 phase still remains magnetically disordered. Finally at $T$ = $T_{N2}$ ($<$ $T_{N1}$), C type AFM (C-AFM) transition of OR2 phase occurs. The ground state of PSMO is constructed by arrays of two coexisting C-AFM phases associated with  OR1 and OR2 respectively.

\par
Our investigation based on the dc and ac susceptibilities, relaxation, aging, and memory measurements reveals the presence of an unusual glassy magnetic phase in the ground state of PSMO even in absence of any obvious microscopic  FM clusters. 
 
\par
Polycrystalline sample of PSMO was prepared following conventional solid state reaction route as described elsewhere~\cite{cha}. The sample was characterized using x-ray powder diffraction (Cu-K$_{\alpha}$), and it was found to be single phase with tetragonal crystal structure (space group: $I4/mmm$; lattice parameters: $a$ = 3.78 \AA, and $c$ = 12.21 \AA) Magnetization ($M$) and ac susceptibility measurements were performed  in a Quantum Design SQUID magnetometer. 

\par
Figure 1(a) depicts the $T$ variation of dc magnetic susceptibility ($\chi = M/H$, $H$ being applied magnetic field.) between 3 K and 320 K measured in zero field cooled heating (ZFCH), field cooling (FC), and field cooled heating (FCH) protocols with $H$ = 500 Oe. All the features as observed in the $\chi(T)$ data nicely follow the magnetic/structural phase diagram of Pr$_{1-x}$Sr$_{1+x}$MnO$_4$ as proposed before~\cite{nor}. Near 300 K, ZFCH curve shows a sharp rise followed by a broad peak around $T_{N1}$ = 200 K.  It is to be noted that $\chi$ above $T_{N1}$ does not follow Curie-Weiss behaviour indicating that the magnetic phase above $T_{N1}$ is not strictly paramagnetic. Presumably, it corresponds to a state with short range magnetic correlations.  On cooling below 200 K, ZFCH curve  decreases sluggishly with $T$ and shows another anomaly near $T_{N2}$ = 125 K which is  related to the C-AFM ordering of the OR2 phase. Inset of figure 1(a) displays $\chi(T)$ in ZFCH condition for $H$ = 100 Oe, where these two magnetic transitions are very prominently visible. A clear bifurcation is present between ZFCH and FCH curve below $T_{irr}$ = 250 K.  FC and FCH curves also contain both the anomalies at $T_{N1}$ and $T_{N2}$ similar to ZFCH data. $\chi(T)$ shows a rise with decreasing $T$ below 35 K, which is particularly prominent in the field-cooled data. Thermal hysteresis is present between  FC and FCH curves  between 40  and 250 K which is the  signature of a first order phase transition (FOPT) related to the structural transition from high-$T$ TG to low-$T$ OR phases. 
\par
Figure 1(b) displays isothermal $M$-$H$ curves measured at three different temperatures (6 K, 200 K, and 320 K). $M$($H$) data at 6 K and 200 K show almost linear $H$ dependence as expected from antiferromagnetically correlated spins. However, a closer look at the low-$H$ region (see inset of figure 2(b)) of these two curves reveals the presence of nonlinearity. Particularly, hysteretic behaviour can be clearly seen in the 6 K data. Such feature can be related to the spin canting in the AFM state. Alternatively, it can also be an indication of a glassy magnetic phase~\cite{bin,lue}. Notably, $M(H)$ data at 320 K is perfectly linear in the entire $H$ regime. 

\par
In order to probe the spin dynamics of the low temperature orbitally ordered phase, we measured $M$ as a function of time ($t$) at different constant $T$ well below the thermal hysteresis region. In figure 2(a), $M$ has been depicted in a normalized form $M(t)$/$M$(0), where $M$(0) is the value of $M$ at the start of the relaxation measurement. All the data carry clear signature of magnetic relaxation and it is as large as 1.6\% at 18 K for $t$ = 3600 s. Such large relaxation of $M$ indicates that the magnetic state at low temperature  is associated with some sort of metastability. Interestingly, the magnitude of relaxation decreases with decreasing temperature. This is presumably an indication of thermal effect, where thermal fluctuations associated with the frozen spins diminishes with decreasing $T$.  In addition, the system consists of structural domains (OR1 and OR2), and the number of such domains  (and hence the number of interfaces) is found to increase with increasing $T$~\cite{nor2}.  Therefore, the enhanced relaxation with $T$ can also be related to the increasing number of such domains.   The relaxation data can be best fitted with a modified stretched exponential function given by~\cite{myd,rvc,phi,fre},

\begin{equation}
M(t) = M_i - M_{r}exp[-(t/\tau)^\beta]
\label{expo} 
\end{equation}

This particular relation is widely used to describe the relaxation behaviour of glassy magnetic systems. Here, $M_i$ corresponds to the initial magnetization, $M_{r}$ is the contribution from a glassy part, $\tau$ is the time constant and the exponent  $0 \leq \beta \leq 1$ is linked with the distribution of energy barriers among metastable states. For an ordered FM system, the numbers of local energy minima shrinks to a single global energy minimum and $\beta$ becomes unity. In PSMO, the fitted values of $\beta$ for all $T$s remain between $\sim$0.55 to $\sim$0.6 which falls within the range of $\beta$ values reported earlier for different spin glass (SG) like systems~\cite{wan,bha}.     

\par
The analysis of relaxation data indicates that PSMO has a metastable ground state with a distribution of local energy minima. Such  energy landscape can correspond to  either a disordered magnetic state or a SG  like phase. Simple relaxation measurement can not conclusively distinguish between them. In order to elucidate the exact magnetic  state of PSMO, aging effect was studied carefully. The sample was zero field cooled down to 6 K  followed by an isothermal aging at $H$ = 0 for particular wait time $t_w$. After that $H$ was increased to 100 Oe and $M$ was measured as a function of $t$. The experiment was repeated for different values of $t_w$, namely 1500 s, 3500 s and 5500 s.  Using the resulting wait time dependent $M(t)$ data, it is possible to calculate the relaxation rate/magnetic viscosity $S(t)$ = $\frac{1}{H}\frac{\partial M}{\partial (lnt)}$~\cite{myd}. For an SG, $S(t)$ shows a peak at $t_p\approx t_w$.  Figure 2(b) describes $S(t)$ graph for three different  values of $t_w$s as mentioned before. Evidently, $S$($t$) curves corresponding to $t_w$ = 1500 s, 3500 s, and 5500 s show their peaks at $t_p$ $\sim$ 1100 s, 2000 s and 2500 s respectively. Such systematic shift in peak position with varying $t_w$ is a signature of aging and regarded as a strong evidence for the presence of glassy magnetic phase. The obtained values of $t_p$ do not match exactly with $t_w$, and they are always smaller than the actual wait time. Similar effect was previously observed in glassy magnetic systems where finite spin clusters rather than individual atomic spins are associated with the spin freezing phenomenon~\cite{hari, markovich}.

\begin{figure*}[t]
\vskip 0.4 cm 
\centering
\includegraphics[width = 12 cm]{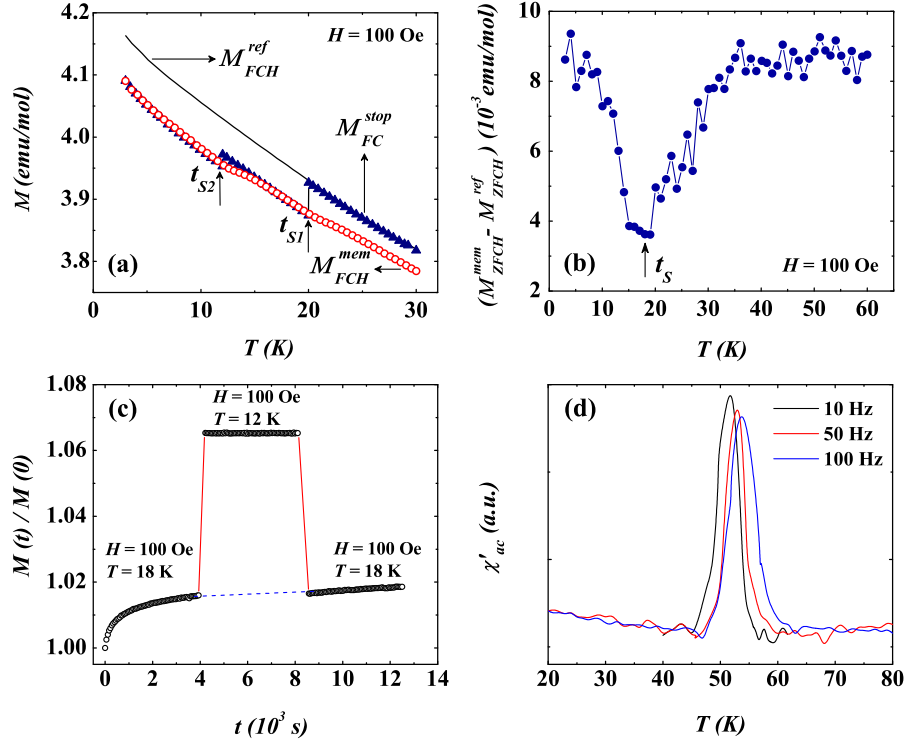}
\caption {(a) depicts field stop memory measurement for PSMO with applied magnetic field of  100 Oe.  $M_{FC}^{stop}$ denotes the cooling curve in 100 Oe with stops at 18 K and 12 K for 3600 s each. During the stop, the magnetic field was reduced to zero. $M^{mem}_{FCH}$ corresponds to the curve during continuous heating maintaining 100 Oe field after the sample being cooled down to base temperature with stops at 18 and 12 K. $M^{ref}_{FCH}$ denotes   the reference FCH curve after the field cooling without any stop. (b) shows the difference curve ($M^{mem}_{ZFCH}$ - $M^{ref}_{ZFCH}$)  as obtained from the ZFC memory measurement. In this measurement protocol,  sample was initially zero field cooled to 3 K with a stop of 10800 s  at 18 K and then continuously heated back to 60 K in presence of 100 Oe field ( denoted as $M^{mem}_{ZFCH}$). Where as, $M^{ref}_{ZFCH}$ is the magnetization data during heating just after the zero field cooling of the sample without any stop.  (c)  Magnetic relaxation data at $H$ = 100 Oe in the zero-field-cooled state measured at 18 K  along with  an intermediate measurement at 12 K. (d) shows temperature variation  of the real part of the ac susceptibility data measured at different frequencies.}
\end{figure*}

\begin{figure}[t]
\vskip 0.1 cm
\centering
\includegraphics[width = 8cm]{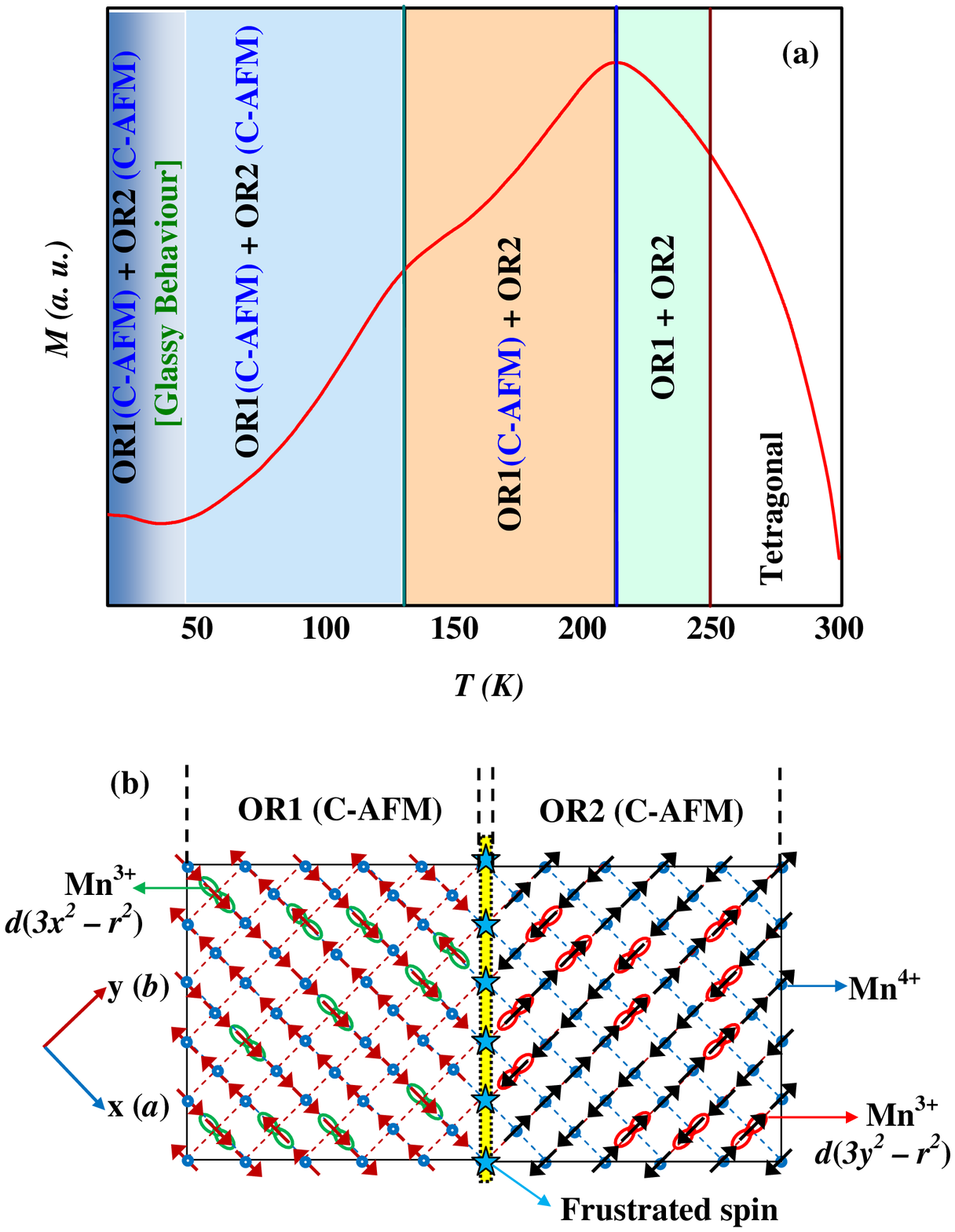}
\caption {(a) Different structural and magnetic phases of PSMO as a function of temperature. The figure also shows the zero-field-cooled magnetization data measured at 100 Oe. (b) shows orthorhombic structural variants OR1 and OR2 along with the spin and orbital arrangements in the ground state. Here $x$ and $y$ orbital axes correspond to the $a$ and $b$ crystallographic axes of the sample respectively.}
\end{figure}

\par
Figure 3(a) depicts the {\it field stop}  experiment based on $M$($T$) measurement  in the FC mode~\cite{jon,sun,sas}. The sample was cooled down to 3 K in 100 Oe with intermediate stops ($H$ being reduced to 0 at the stops) of duration 1 h each at 18 K and 12 K (curve $M^{stop}_{FC}$). Subsequently, the sample was heated back to 30 K and $M$ was measured as a function of $T$ (curve $M^{mem}_{FCH}$). Clearly, while heating  $M$ shows distinct anomaly (marked at $t_{S1}$ and $t_{S2}$) exactly at the same temperatures where the sample was allowed to age for 1 h during cooling. Those  anomalies are clearly absent in the reference field-cooled heating data ($M^{ref}_{FCH}$), which was recorded without any stop while cooling. Such signature of magnetic history is a convincing evidence for the glassy magnetic state. However, non-interacting  superparamagnetic nanoparticles can show similar field stop memory effect originating from the distribution of magnetic relaxation time among the particles~\cite{sas2,ban}. To rule out such possibility in PSMO, we measured magnetic memory in the ZFC condition, which has the same protocol as that field stop memory experiment baring the fact that sample is cooled in $H$ = 0 with intermediate stop at 18 K. We have shown the ZFC memory curve in fig. 3 (b), where the reference magnetization curve was subtracted out from the original memory curve. It shows a clear dip  at 18 K (marked by $t_S$ in the figure), exactly the temperature where sample was allowed to relax for 3 h during cooling. This conclusively rules out the  possible origin of memory from noninteracting superparamagnetic nanoparticles.

\par
The observed glassy phase is further supported by the $t$ dependent memory  measurements (fig. 3 (c)). The sample was allowed to relax in 100 Oe field at 18 K in the ZFC condition. After 3600 s, $T$ was decreased to 12 K without altering $H$ and the relaxation was recorded for another 1 h.  In the last segment of this $M$($t$) measurement, sample was again heated back to 18 K without changing the magnetic field. Interestingly the system starts to relax from a point where it ended up in the first segment. It means that the system can remember its earlier state in spite of such negative $T$ cycling, which is a clear evidence for the SG like ground state of PSMO.  Such observations of memory is in line with the  Hierarchical and Droplet Model  description of spin glasses~\cite{jon, sas, sas2, mat, mat2, mat3, ajb}.

\par
It is now pertinent to investigate the onset temperature for the SG freezing. We performed magnetic memory experiment (both temperature and time dependent) at various $T$ range.  The field stop experiments on $M(T)$ curve (similar to as described in fig. 3 (a)) fail to show any signature of memory above about 50 K.  The $t$ dependent memory measurements (similar to as described in fig. 3 (c)) also rules out the possibility of SG-like phase above about 50 K. This is also supported by the $T$ variation of ac susceptibility ($\chi_{ac}$) measurement on the sample at different frequencies. The real part of $\chi_{ac}$ ($\chi_{ac}^{\prime}$) shows a clear peak around  50 K (see fig. 3 (d)), and the peak  shifts to higher $T$ with increasing frequency.  It is to be noted that no other prominent anomaly in $\chi_{ac}$ was observed in the entire temperature range (5-300 K) of measurement (curve not shown here).

\par
Among single layered manganites, glassy magnetic state was previously reported in case of Eu$_{0.5}$Sr$_{1.5}$MnO$_4$, La$_{1.1}$Sr$_{0.9}$MnO$_4$ and Pr$_{1-x}$Ca$_{1+x}$MnO$_4$ (0.35 $\leq x \leq$ 0.5)~\cite{yu,mat,mat2,mat3}. Unlike AMnO$_3$ type perovskite manganites, long range FM state is absent in single layered system, and therefore a relatively easy picture of competing interactions between FM and AFM clusters leading to a glassy phase cannot be invoked. The glassy magnetic phase in the above mentioned single layered manganites was explained on the basis of the fragmentation of charge-orbital ordered state down to nanometer scale. This leads to  the mixture of AFM and FM bonds  in case of   CE-AFM structure of those SLMs resulting competing interaction.  It was found that such competing magnetic interaction in the nanometer scale can only exist in underdoped  and optimally doped ($x \leq$ 0.5)  single layered sample of general formula T$^{3+}_{1-x}$D$^{2+}_{1+x}$Mn$^{3+}_{1-x}$Mn$^{4+}_{x}$O$^{2-}_4$ (T and D are respectively trivalent and divalent cations). For $x >$ 0.5, the SG state disappears due to the presence of extra holes~\cite{mat3}.

\par          
The presently studied sample is actually an overdoped composition ($x$ = 0.78), and therefore the above model of SG arising from the nanoscale disorder is not applicable here. The most likely scenario for such glassy phase is the existence of different orthorhombic variants with dissimilar spin structures. In fig. 4 (a), we have depicted different phases of PSMO as obtained from  our magnetic measurements and previous reports.  The transition from TG to OR phase takes place over a wide temperature range (approximately 250 K to 40 K), which is characterized by the thermal hysteresis in $M$ as depicted in fig. 1. The most important observation is that both the OR variants have C-type antiferomagnetically ordered state at least below 125 K, and these magnetically ordered variants coexist with each other down to the lowest $T$.

\par
In fig. 4 (b), we have shown two such variants along with their spin and orbital arrangements in the MnO$_2$ layer. The $e_g$ orbitals are shown for Mn$^{3+}$ ions only where they are occupied by electrons.  The $e_g$ orbitals are aligned either along $x$ (for $d({3x^2-r^2})$ type OO) or $y$ (for $d({3y^2-r^2})$ type OO) direction and their orientations are mutually orthogonal in two variants.  The spin structure in PSMO strongly depends upon the nature of the orbital ordering. The ordered orbitals along the $x$ or $y$ direction favours the electronic transfer leading to FM double exchange interaction only along the respective directions. This along with the Mn-O-Mn type superexchange give rise to  C-type antiferromagnetism, where we get antiferromagnetically coupled ferromagnetic chains along the $x$ or $y$ direction within a single MnO$_2$ layer.   

\par
The FM exchange interaction, which originates from the electron transfer along the ordered orbitals, will  have direction  same as the axis of orbital ordering  (here either $x$ or $y$). This will in its turn favour the spins to lie along the $x$ or $y$ direction. As a result, similar to the OO state, two orthorhombic variants will also have mutually perpendicular spin anisotropy axis (see fig. 4 (b)). 

\par            
The FM and AFM bonds in a homogeneous C-type AFM phase do not introduce magnetic frustration. However, situation can be quite different at the interface of the two orthorhombic variants with mutually perpendicular spin anisotropy axis. As shown in fig. 4 (b), such spin arrangements can introduce frustration at the interface. If one looks carefully, the spins in a chain are ferromagnetically coupled along the $x$ axis in the left variant (OR1), however, in the right variant (OR2) spins are antiferromagnetically coupled along the same $x$ direction. Therefore, the variants can introduce competing magnetic interaction in the interfacial region. Such competition along with disorder or defects due to doping can give rise to  spin freezing. It was already predicted theoretically  that anisotropy can indeed play a key role for the development of SG state~\cite{ajb}. The average size of the variants in PSMO was reported to be about few hundred nanometers, and so at low temperature there will be enough interfacial frustrated spins for the SG like states to evolve. In the above discussed model, we have considered the interface to be perfectly sharp. However, in reality it may be rough and it can introduce additional randomness in spin structure. The above scenario is quite similar to the cluster glass in perovskite manganites and cobaltites with coexisting FM and AFM domains~\cite{ito}. The only difference is that here the coexisting structural domains are all AFM but with unequal spin anisotropy axis. 

\par
It is evident from fig. 4  that magnetic frustration in the system can actually exist below 125 K where both  OR1 and OR2 become antiferromagnetically ordered. However, we failed to see characteristic features (such as memory in the $t$ and $T$ dependent $M$) of a spin glass above  about 50 K.  $\chi_{ac}$ also shows clear frequency dependent anomaly around 50 K signifying the onset of spin freezing. Therefore, despite the prevailing spin frustration, the system only shows SG phase below about 50 K.  This might be related to the thermal fluctuations of spin, which only become weak enough below 50 K for spin freezing to occur.  However, it is to be noted that magnetic relaxation is  present even temperatures as high as $\sim$ 200 K, and it completely vanishes only above the region of hysteresis. This observed relaxation at higher $T$  is presumably not connected to the glassy magnetic phase or frustration, rather it corresponds to the landscape of metastable states associated with the disorder driven FOPT~\cite{chadda}. The time dependent nucleation of a particular phase within the region of FOPT can also give rise to slow dynamics of $M$  provided the coexisting phases have distinctive magnetic character.       

\par
In conclusion, we observed glassy magnetic state in single layered manganite PSMO. Due to the quasi 2D structure, SLMs do not sustain metallic FM phase, and therefore the coexistence of FM/AFM clusters can not be the cause of the glassy state. The likely origin is the existence of structural domains with C-type antiferromagnetic ordering,  where spin alignment is  closely related to the nature of the orbital ordering in two adjacent structural domains. The interface between two domains can give rise to spin frustration leading to a SG like state. Such scenario can be important in understanding glassy magnetic state in several magnetic oxides.    

\par
S. C. wishes to acknowledge Council of Scientific and Industrial Research, India.

\end{document}